\documentclass[useAMS,usenatbib]{mn2e}

\usepackage{graphicx}
\usepackage{dcolumn}
\usepackage{bm}
\usepackage{amsmath}
\usepackage{amssymb}
\usepackage{url}

\title[GRB Detection Rate with CTA]
{Prospects for Detecting Gamma-Ray Bursts at Very High Energies with the Cherenkov Telescope Array}
\author[J. Kakuwa et al.]
{
Jun~Kakuwa,$^{1}$\thanks
{E-mail: kakuwa@theo.phys.sci.hiroshima-u.ac.jp (JK)}
Kohta~Murase,$^{2}$
Kenji~Toma,$^{3}$
Susumu~Inoue,$^{4}$
Ryo~Yamazaki,$^{5}$
\newauthor
and
Kunihito~Ioka$^{6}$
\\
$^{1}$Department of Physical Science, Hiroshima University,
      Higashi-hiroshima 739-8526, Japan\\
$^{2}$Center for Cosmology and AstroParticle Physics, Ohio State University,
      191 West Woodruff Avenue, Columbus, OH 43210, USA\\
$^{3}$Department of Earth and Space Science, Osaka University,
      Osaka 560-0043, Japan\\
$^{4}$Institute for Cosmic Ray Research, University of Tokyo, Kashiwa, 
      Chiba 277-8582, Japan\\
$^{5}$Department of Physics and Mathematics, Aoyama Gakuin University,
      Sagamihara 252-5258, Japan\\
$^{6}$KEK Theory Center and the Graduate University for Advanced Studies (Sokendai),
      Tsukuba 305-0801, Japan\\
}
\begin{document}

\date{Accepted 2012 June 11. Received 2012 June 07; in original form 2011 December 27}

\pagerange{\pageref{firstpage}--\pageref{lastpage}} \pubyear{????}

\maketitle

\label{firstpage}

\begin{abstract}
We discuss the prospects for the detection of gamma-ray bursts (GRBs) by the Cherenkov Telescope Array (CTA), the next generation, ground-based facility of imaging atmospheric Cherenkov telescopes (IACTs) operating above a few tens of GeV. By virtue of its fast slewing capabilities, the lower energy threshold compared to current IACTs, and the much larger effective area compared to satellite instruments, CTA can measure the spectra and variability of GRBs with excellent photon statistics at multi-GeV energies, which would revolutionize our understanding of the physics of GRBs, test their validity as the origin of ultra-high-energy cosmic rays, and provide powerful probes of the extragalactic background light as well as Lorentz-invariance violation. Employing a model of the GRB population whose properties are broadly consistent with observations by the Gamma-ray Burst Monitor (GBM) and Large Area Telescope (LAT) onboard {\it Fermi}, we simulate follow-up observations of GRBs with the Large Size Telescopes (LSTs), the component of CTA with the fastest slew speed and the best sensitivity at energies below a few hundred GeV. For our fiducial assumptions, we foresee that the LSTs can detect $\sim$~0.1 GRBs per year during the prompt phase and $\sim 0.5$ per year in the afterglow phase, considering only one array site and both GBM and the Space-based multi-band astronomical Variable Object Monitor ({\it SVOM}) as the alert instruments. The detection rates can be enhanced by a factor of about 5 and 6 for the prompt emission and the afterglow, respectively, assuming two array sites with the same sensitivity and that the GBM localization error can be reduced to less than 1$^\circ$. The expected distribution of redshift and photon counts are presented, showing that despite the modest event rate, hundreds or more multi-GeV photons can be anticipated from a single burst once they are detected. We also study how the detection rate depends on the intrinsic GRB properties and the delay time between the burst trigger and the follow-up observation. 
\end{abstract}


\section{Introduction}

Gamma-ray bursts (GRBs) are the most violent explosive phenomena in the Universe.  
The prompt emission of GRBs typically has luminosity of $L_{\gamma} \sim 10^{51}$--$10^{52}~{\rm erg}~{\rm s}^{-1}$ at $\sim$~0.1--1~MeV, and the following long-lasting afterglow emission is observed in the radio to X-ray bands~\citep[see reviews,][]{Mes06,Zha07}. 
High-energy gamma-ray emission above $\sim 0.1$~GeV was first seen by the Energetic Gamma-Ray Experiment Telescope (EGRET) detector on-board the Compton Gamma Ray Observatory~\citep[e.g.,][]{Hur+94,Gon+03}.  
The Large Area Telescope (LAT) on-board the \textit{Fermi} satellite has recently detected high-energy gamma rays from a much larger sample of GRBs~\citep{Abd+09a,Abd+09b,Abd+09c,Abd+10,Abd+11,Ack+10,Ackermann2011}. 
The LAT GRBs exhibit the following features~\citep[see recent reviews,][]{Granot2010,BG11,Pee11,Ino+12}.
(a) Most of the LAT GRBs do not show significant suppression at the high-energy end of their spectra, though observed LAT limits on the GeV fluence for the GRBs detected by the Gamma-ray Burst Monitor (GBM) onboard the {\it Fermi} satellite may suggest a steeping or cutoff in the high-energy spectrum~\citep{Beniamini2011};   
(b) Some of the LAT GRBs have an anomalous extra component in the $>0.1$~GeV range~\citep{Abd+09c,Ack+10,Ackermann2011}, while others show a high-energy spectrum consistent with the Band function~\citep{Band1993};
(c) In many cases, the emission onset in the $>0.1$~GeV energy range is delayed relative to that in the $<1$~MeV energy range.
(d) The LAT GRBs often show the long-lived high-energy emission lasting longer than the duration of the sub-MeV component; 
(e) Not only long GRBs but also short GRBs seem to have the above features~\citep{Ack+10}. 
Understanding these features is likely to give us important clues to GRB mechanisms and related astrophysics. 

The Cherenkov Telescope Array (CTA), under plan as the next generation international gamma-ray observatory~\citep{CTA10,CTA11}, will provide a great step forward in studying these issues. 
CTA will be constructed at two sites, one each in the northern and southern hemispheres, and comprises three types of telescopes: 
the Large Size Telescopes (LSTs) with 23~m diameter and 4.6$^\circ$ field of view (FOV); 
the Medium Size Telescopes (MSTs) of 12~m and 8$^\circ$ FOV; and the Small Size Telescopes (SSTs) of 7~m and 10$^\circ$ FOV. 
With CTA, the sensitivity will improve by a factor of 5--10 in the 0.1--10~TeV range compared to existing Cherenkov telescopes. 
It will cover about 5 decades in energy, including energies below a few tens of GeV and above $100$~TeV, and the angular and energy resolution will be appreciably increased. 
In the following, we summarize open issues that can be unraveled by CTA.
We refer the reader to \citet{Ino+12} for an extensive overview on the science prospects for GRB observations with CTA.
\begin{enumerate}
\item
The bulk Lorentz factor $\Gamma$ is one of the key quantities in understanding the properties of relativistic jets making GRBs.  It is thought to be limited to order of $\sim 1000$ in the classical fireball model with baryons, whereas higher Lorentz factors may be achieved by Poynting-dominated jets~\citep[e.g.,][]{SDD01} or radiation-dominated jets with dissipation via e.g., jet-confinement~\citep{Iok+11} or baryon-entrainment~\citep{Iok10}.  The bulk Lorentz factor can be constrained from observations of the high-energy end of the spectra~\citep{LS01}, and absence of high-energy cutoffs for the LAT GRBs have indeed given us lower limits on $\Gamma$~\citep[e.g.,][]{Abd+09a,Abd+09b,Abd+09d}.  
In the case of GRB 090926A~\citep{Ackermann2011}, a sharp softening of the spectrum is observed at $\sim 1.4$~GeV, and interpreting this as the pair-creation cutoff leads to $\Gamma \approx 720$.  
However, such estimates can be affected by the finite extent of the emission region~\citep{Bar06}, time-dependence of the photon field~\citep{Gra+08,AM11} and multiple emission regions~\citep{Li10,Aoi+10,Has+11}, so that the resulting high-energy spectra can be more complicated.  For more detailed studies, observations with much better statistics are required.  
CTA will be ideal for these purposes.  
\item
The mechanism of the prompt emission is one of the most critical issues in the theory of GRBs. 
In the classical scenario, $\sim$~MeV emission is explained by the optically-thin synchrotron radiation from electrons accelerated at internal shocks~\citep{RM94}.  
However, this classical scenario has several problems in explaining observations~\citep[see reviews,][]{Mes06,Zha07}, and many alternative scenarios such as the photospheric dissipation scenario~\citep[e.g.,][]{Tho94,MR00,RM05,Iok+07,Bel10,Mur+12} and magnetic dissipation scenarios~\citep[e.g.,][]{Lyu06,ZY11,MU11} have been suggested.  For the mechanisms of the $>0.1$~GeV emission, both leptonic and hadronic emission have been have been proposed.  Synchrotron emission~\citep{Wan+09,Iok10,Dai+11}, synchrotron self-Compton (SSC) emission~\citep{CGP10a,CGP10b}, external inverse-Compton (EIC) emission~\citep{TWM09,TWM11}, and proton synchrotron or proton-induced cascade emissions~\citep{AGM09,AIM10,RDF10,Mur+12} are currently viable explanations.  However, once the prompt emission is detected by CTA, it may be possible to discriminate among the emission mechanisms by analyzing the very-high-energy (VHE) spectrum and variability with high photon statistics. 
\item
It is believed that the long-lived high-energy component is related to the afterglow emission.  In the simplest scenario, the high-energy emission is explained by synchrotron emission from the non-radiative external forward shock with extreme parameters~\citep{KB11} or radiative external forward shock~\citep{Ghisellini10}.  The Klein-Nishina effect may play a role in the initial rapid decay~\citep{Wan+10}, but the more natural explanation is the gradual turn-off of the prompt emission~\citep{LW11,He+11,Max+11}. 
One of the important tests for synchrotron external shock scenarios is to see the maximum synchrotron cutoff that decreases as the Lorentz factor declines~\citep{PN10}, and CTA may eventually see the SSC emission component. 

Observations by CTA can also provide a clue to the origin of X-ray shallow-decay emission.  The shallow-decay behavior is observed in most X-ray afterglows of \textit{Swift} GRBs~\citep{Nou+06,Obr+06,Zha+06}, though it was not predicted by the standard afterglow model~\citep{MR97,SPN98}. 
Its origin has been debated for many years, and possible models include energy injection \citep[e.g.,][]{DL98,RM98,Zha+06}, the long-lasting internal activity \citep{Ghisellini07,Kum+08,Mur+11}, the long-lasting reverse shock~\citep{GDM07,UB07}, time-dependent microphysics~\citep{ITYN06}, effect of cosmic ray escape~\citep{Der07}, bulk Compton emission~\citep{Pan08a}, prior emission~\citep{Yam09}, and multi-component jets~\citep[e.g.,][]{EG06,TIYN06}.  
These models should predict different high-energy emissions~\citep{Fan+08,Pan08b,Mur+10,Mur+11}, so that the simultaneous observations by {\it Swift} and the Space-based multi-band astronomical Variable Object Monitor ({\it SVOM})~\citep{Paul2011} as well as CTA would be useful for discriminating among the models.  
On the other hand, high-energy emission has been observed concurrently with X-ray flares, which are often seen in the early afterglow phase~\citep{Abd+11}.  Although they are considered to originate from some internal dissipation processes rather than external shocks~\citep{IKZ05}, their mechanism and the origin of the high-energy emission is still unclear.  CTA may provide a breakthrough for revealing these problems.  
\item
VHE signals from GRBs can be useful for other purposes. 
GRBs may be the main sources of the observed ultrahigh-energy cosmic rays (UHECRs)~\citep{Wax95,Vie95}.  If this hypothesis is true, VHE gamma rays provide one of the crucial probes, and detections of characteristic signals, including hadronic cascade radiation~\citep{BD98,PW05,DA06,AIM09,Mur+12}, ion synchrotron radiation~\citep{BD98,Tot98a,Tot98b}, synchrotron pair-echo emission induced by ultra-high-energy photons~\citep{Mur11}, and emission via nuclear photodisintegration and Bethe-Heitler pair production~\citep{MB10}, are relevant to test this GRB-UHECR hypothesis. 
Also, VHE gamma rays from GRBs are useful as a probe of the Universe.  Since they are attenuated by the extragalactic background light (EBL), one can constrain the EBL by measuring the attenuation as in the case of blazars~\citep[e.g.,][]{Aha+06,Alb+07}.  VHE gamma rays also lead to secondary emission as a result of interactions with the EBL, and detections or non-detections of this pair-echo emission enable us to probe the uncertain intergalactic magnetic fields in voids as well as the EBL~\citep[e.g.,][]{RMZ04,Tak+08,Mur+09}.  
GRBs may occur even at the very distant universe, $z > 5$~\citep[e.g.,][]{Kaw+06,Sal+09,Tan+09}, and VHE gamma rays might be generated by GRBs originating from population III stars~\citep{TSM11}.  If we detect the attenuated VHE gamma rays from such distant GRBs, it would be possible to constrain the EBL in the high-redshift universe~\citep{Gil+09,Ino+10} and/or cosmic magnetic fields at high redshifts~\citep{Tak+11}.  
In addition, one can obtain insight into fundamental physics.  For example, using the temporal (and spectral) information, one can test the Lorentz-invariance violation predicted in some quantum gravity theories~\citep[e.g.,][]{Ame+98,Abd+09d,SM10}.
\end{enumerate}

As described above, CTA should be a powerful tool to understand GRB physics and test other various ideas and hypotheses related to GRBs.  VHE gamma-ray astronomy has now been firmly established by state-of-the-art IACTs such as H.E.S.S., MAGIC, and VERITAS.  A firm detection of GRBs has not been reported so far~\citep{Alb+07,Aha+09,Ale+10}, but CTA may eventually achieve the goal with much lower energy threshold and much better sensitivity, 
which will also allow us to obtain time-resolved spectra of much better quality than ever before in the energy range greater than a few tens of GeV. 
In this work, we report the prospects for detecting long GRBs (durations longer than 2~sec) with CTA, and how the results depend on some of the array performance and GRB properties. 
The situation considered here is that GRBs are detected by some satellites and followed up by the LSTs, which will play a principal role in GRB observations. 
In our simulation, we mainly assume {\it Fermi}/GBM as the burst trigger, which has already provided data with high statistics and may be active in the CTA era. 
We also make a rough estimate for the alerts from the {\it SVOM} satellite, which is planned to be launched before CTA operation. 
Some complementary aspects of this work are also presented in \citet{Ino+12}.

The organization of the paper is as follows. 
In section~2, we describe the intrinsic GRB properties of both prompt and afterglow emissions in our simulations. 
In section~3, we model the performance of the LSTs and the follow-up observation of GRBs.
Performing Monte Carlo simulation, we estimate the detection rate of GRBs with the LSTs in section~4.  Finally, section~5 is devoted to summary and discussion.
We use cosmological parameters $H_0 = 70$~km~s$^{-1}$Mpc$^{-1}$,  $\Omega_{m} = 0.3$ and  $\Omega_\Lambda = 0.7$.

\section{Intrinsic GRB properties}
\label{sec:GRB_properties}

In this section, we describe our method of generating the intrinsic GRB samples in our Monte Carlo simulations and the assumed EBL model. 
It provides each sample with the prompt and afterglow properties using a luminosity function and well-known spectral correlations. 
We consider the GRB samples only with the duration $>2$~sec and redshift $<5$. 
It is also shown that our sample is consistent with some observational results of GBM and LAT onboard {\it Fermi}.

\subsection{Prompt emission}
\label{sebsec:prompt}


We assume the luminosity function of GRBs, defined as the rate of GRBs per unit 
comoving volume at a redshift $z$ and per logarithmic interval 
of peak luminosity $L_{\rm p}$ for 1--10$^4$~keV in the cosmological rest frame, to be
\begin{equation}
\Psi(L_{\rm p},z) = \rho(z)\phi(L_{\rm p})~~. 
\label{eq:Psi}
\end{equation}
Following \citet{Wanderman2010}, we describe the functional forms of the GRB formation rate 
$\rho(z)$ and the local luminosity function $\phi(L_{\rm p})$ as broken power-laws: 
\begin{eqnarray}
\rho(z) \propto \left\{ 
\begin{array}{ll}
(1+z)^{\alpha_z} &  z<z_p  \\
(1+z)^{\beta_z} &  z>z_p  \\
\end{array} 
\right.~~, 
\label{eq:rho_z}
\end{eqnarray} 
and 
\begin{eqnarray}
\phi(L_{\rm p}) \propto \left\{ \begin{array}{ll}
L_{\rm p}^{\alpha_L} &  L_{\rm p}<L_*  \\
L_{\rm p}^{\beta_L} &  L_{\rm p}>L_*  \\
\end{array} \right.~~. 
\label{eq:phi_L}
\end{eqnarray} 
In our fiducial case, $\alpha_z = 2.1^{+0.5}_{-0.6}$, $\beta_z = -1.4^{+2.4}_{-1.0}$, $z_p = 3.1^{+0.6}_{-0.8}$, $\alpha_L = -0.17^{+0.1}_{-0.2}$, $\beta_L =-1.44^{+0.6}_{-0.3}$ and $L_* = 10^{52.5\pm 0.2}$ erg~s$^{-1}$ are adopted~\citep{Wanderman2010}.
The normalization is determined so that we reproduce the observed trigger rate for GBM of 250~yr$^{-1}$ assuming its trigger condition as the peak photon flux of $1.5$~ph~s$^{-1}$cm$^{-2}$ in the 8--10$^3$~keV band, satisfied by $\simeq 90$~\% of actual GBM bursts. 
In the following, for given values of $L_{\rm p}$ and $z$, we generate parameters of a burst regarding the prompt and afterglow emission.


In order to model the prompt GRB observation with CTA, the simulated $T_{90}$ distribution must be a  good approximation of the observed one, where $T_{90}$ is the time needed to accumulate from 5~\% to 95~\% of observed photon counts. 
The reason is that the delay time $T_{\rm delay}$, which is the start time of a follow-up observation as measured from the burst trigger time, is required to be shorter than the burst duration for the detection of the prompt emission. 
In order to determine $T_{90}$ of our simulated GRB samples, at first we assume that the duration of the prompt emission 
in the energy range where the LSTs are most sensitive (from $\sim 10$~GeV to $\sim 100$~GeV) is equal to $T_{90}$ in the GBM band, and that the prompt phase provides constant gamma-ray luminosity $L_{\rm ave}$, evaluated by averaging the highly variable light curve. 
To determine the value $L_{\rm ave}$ from the peak luminosity $L_{\rm p}$, we take the average of $({E}_{\rm iso}/T_{90}')/L_{\rm p}$ in the observed events~\citep{Ghirlanda2004,Ghirlanda2009,Ghirlanda2010}, where $T_{90}'=T_{90}/(1+z)$ and ${E}_{\rm iso}$ is the isotropic-equivalent gamma-ray energy. 
The data of $T_{90}$ are taken from the GCN circulars. 
Then, we found $(E_{\rm iso}/T_{90}')/L_{\rm p}=0.3\pm0.2$. 
Despite the large scatter, we set 
\begin{equation}
L_{\rm ave}= 0.3L_{\rm p}~~, 
\end{equation}
for all the simulated bursts. 
Next, we examine a correlation between $\log L_{\rm p}$ and $\log E_{\rm iso}$~\citep{Ghirlanda2012} employing a chi-square fit with the effective variance weighting, and obtain 
\begin{equation}
\log E_{\rm iso,52}=1.1\log L_{\rm p,52}+0.56~~, 
\end{equation}
with a standard deviation of 0.49~dex. 
Using this relation and taking the standard deviation into account, $E_{\rm iso}$ is determined by $L_{\rm p}$ in our simulation. 
Then, we determine the duration as 
\begin{equation}
T_{90}'=E_{\rm iso}/L_{\rm ave}~~, 
\end{equation}
and $T_{90}=(1+z)T_{90}'$.

As the intrinsic spectral shape of the GRB prompt emission, we assume the Band function, which is given by~\cite{Band1993} as 
\begin{eqnarray}
N_E \propto \left\{ \begin{array}{ll}
E^{\alpha} \mathrm{exp}(-E/E_0) &  E \leq (\alpha - 
\beta)E_0  \\
E^\beta &  E \geq (\alpha - \beta)E_0  \\
\end{array} \right.~~, 
\end{eqnarray} 
where $\alpha$ and $\beta$ are the high- and low-energy photon indices, respectively. 
If $\beta < -2$ and $\alpha > -2$, then the $\nu F_\nu$ spectrum has a peak at $E_{\rm p} = (2+\alpha)E_0$.
For given $L_{\rm p}$, we determine $E_{\rm p}$ according to the $E_{\rm p}$--$L_{\rm p}$ relation examined by \cite{Ghirlanda2009}~\citep[the so called Yonetoku relation:][]{Yonetoku2004} taking the standard deviation into account. 
The average luminosity $L_{\rm ave}$ determines the normalization of the Band function. 
The photon indices $\alpha$ and $\beta$ are determined according to the distribution that was actually observed for bright BATSE bursts \citep{Kaneko2006}. 
In our fiducial case, only the bursts with $\beta<-2$ are treated. 

In summary, the luminosity function provides $L_{\rm p}$ and $z$ for each generated burst, which are subsequently related to $E_{\rm iso}$, $L_{\rm ave}$, $T_{90}$ and ${E_{\rm p}}$. 
Photon indices $\alpha$ and $\beta$ are determined independently.


For a check of the validity of our method that generates GRB samples, we simulate the bursts triggering GBM and compare their $T_{90}$ and fluence distributions with observations. 
The observed data (from GRB~080714 to 101130) were taken from the GCN circulars. 
In the top panel of Figure \ref{fig:S_vs_T90,S-pdf}, the scatter plot of the fluence and $T_{90}$ for both the simulated and the observed GBM samples are shown. 
In the bottom panel of the same figure, we show the simulated fluence distribution for GBM bursts in the 8--10$^3$~keV band superposed on the observed one. 
Both panels of Figure~\ref{fig:S_vs_T90,S-pdf} show that the actually observed properties of fluence and $T_{90}$ are well reproduced by our model. 
We find that $20~\%$ of the GRBs in our simulation have $T_{90}$ larger than 100 sec, the typical value of $T_{\rm delay}$ assumed in our fiducial case as described later. 
This is only slightly larger than that for actually detected events ($16~\%$). 
Our modelled duration distribution can be considered to be in sufficiently good agreement with the observed one for our purposes of estimating the detection rates. 
\begin{figure}
\centerline{\includegraphics[width=1\columnwidth]{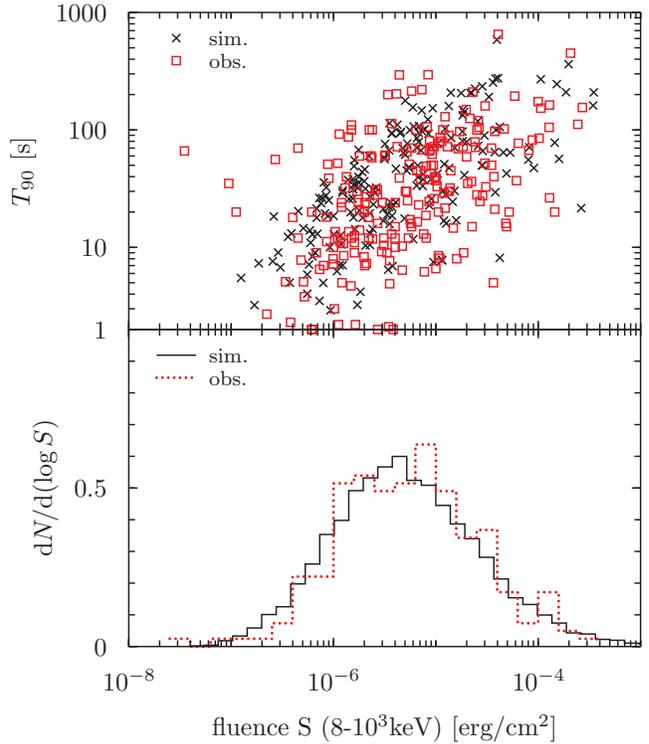}}
\caption{
{\it Top panel}: The scatter plot of fluence and $T_{90}$ for the simulated (black cross) and the observed (red square) GBM samples. 
{\it Bottom panel}: The simulated fluence $S$~(8--10$^3$~keV) distribution in the form of ${\rm d}N/{\rm d}(\log S)$ for GBM bursts (black solid line) superposed on the observed one (red dotted line), which is normalized as $N=1$. 
}
\label{fig:S_vs_T90,S-pdf}
\end{figure}


Furthermore, we verify that our method roughly agrees with the {\it Fermi}/LAT detection rate, which is estimated as follows. 
First, we estimate the all-sky event rate to be $\sim1000$~yr$^{-1}$ for the bursts satisfying $T_{\rm 90}>2$~sec and $z<5$ by deconvolving the observed GRB trigger rate of 250~yr$^{-1}$ with the FOV of GBM (9.5~sr)\footnote{http://gammaray.msfc.nasa.gov/gbm/instrument/description\\ /character.html.} and the trigger efficiency of GBM obtained with our simulation. 
Next, we calculate the detection efficiency of LAT for the bursts satisfying both $T_{\rm 90}>2$~sec and $z<5$ using our intrinsic GRB model, where a burst is judged to be detectable with LAT when the number of detected photons $> 100$~MeV by LAT is larger than 10. 
Here we take the LAT effective area from~\cite{Atwood2009} and consider the signal-dominated regime where the background can be neglected, and take into account its dependence on incidence angle $\theta$ simply by a factor of $\cos\theta$. 
The incidence angle is isotropically distributed in the LAT FOV. 
Then, combining the above derived values with the LAT FOV of 2.4~sr~\citep{Atwood2009}, we estimate the LAT detection rate to be 12~yr$^{-1}$, which is roughly consistent with the observed one, about 7--8~yr$^{-1}$ \citep{Granot2010}. 
Strictly speaking, our simulated LAT detection rate is slightly overestimated. 
Better agreement may possibly be achieved by accounting for spectral softening below the LAT band suggested for subset of events~\citep{Beniamini2011}, more sophisticated model of LAT detection conditions, etc. 
We leave this issue as future work (though a brief discussion will be given in Section \ref{sec:summary}). 


For the estimate of the detection rate with CTA/LSTs, we also simulate the case in which the prompt emission has an extra hard component in addition to the Band spectral component. 
Such an extra component has been confirmed by {\it Fermi}/LAT in some bursts \citep{Abd+09b,Ack+10}. 
However, at present, the properties of the extra component, such as its spectral slope, amount of released energy, and the fraction of events with the extra component, are highly uncertain. 
Therefore we introduce another parameter, $R_{\rm extra}=L_{\rm extra}/L_{\rm ave}$, where $L_{\rm extra}$ is the luminosity of the extra component in the 0.1--100~GeV band. 
Roughly speaking, $R_{\rm extra}$ is the ratio of the LAT band luminosity to that of GBM. 
For LAT bursts, $R_{\rm extra}$ is about a few $10$~\%~\citep{Ghisellini10}, and from the EGRET and LAT observations, the photon index in the GeV region is about $-2$ on average~\citep[][]{Dingus1995,Ghisellini10}. 
Below, for simplicity, we assume that all simulated bursts have the same $R_{\rm extra}=0.1$ and a photon index of $-2$. 
\begin{figure}
\centerline{\includegraphics[width=1\columnwidth]{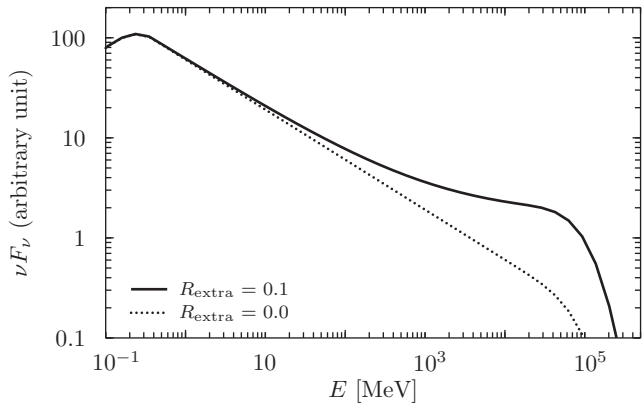}}
\caption{
Examples of the $\nu F_\nu$ spectra of the prompt emission for the case of $R_{\rm extra}=0$ (dotted curve) and 0.1 (solid curve) are illustrated. 
For both cases, $\alpha = -1.0$, $\beta = -2.5$, $E_{\rm p} = 250$~keV, and $z=1.0$ are assumed. 
Cutoffs near 100~GeV are caused by the EBL attenuation modeled by~\citet{Razzaque2009}. 
}
\label{fig:sample_spec.eps}
\end{figure}
Figure~\ref{fig:sample_spec.eps} illustrates examples of the $\nu F_{\nu}$ spectrum of the prompt emission for the case of $R_{\rm extra}=0$ and 0.1. 
In both the cases, $\alpha = -1.0$, $\beta = -2.5$, $E_{\rm p} = 250$~keV, and $z=1.0$ are adopted.  Cutoffs near 100~GeV are caused by the EBL attenuation modeled by \citet{Razzaque2009}. 
We assume this EBL model in our fiducial case (see Section \ref{subsec:EBL}). 

\subsection{Afterglow emission}


High-energy afterglow emission in the CTA band is also highly uncertain. 
Then, following \citet{Ghisellini10}, we assume a simple phenomenological model to describe the luminosity of the afterglow emission in the 0.1--100~GeV band in the cosmic rest frame as 
\begin{equation}
L_{\rm AG}(T') = 10^{52} {\rm erg}~{\rm s}^{-1}
\left(\frac{E_{\rm iso}}{10^{54}{\rm erg}}\right)
\left(\frac{T'}{10~{\rm s}}\right)^{p_t}~~,
\label{eq:afterglowCTA}
 \end{equation}
where $p_t$ is the temporal decay index, and $T'= T/(1+z)$ is the elapsed time from the burst trigger in the cosmic rest frame. 
We choose the normalization to reproduce the result seen in Figure~4 of \citet{Ghisellini10}. 
Now the isotropic-equivalent energy and the burst duration of the prompt emission, $E_{\rm iso}$ and $T_{90}$, are determined as described in the previous subsection. 
We assume $L_{\rm AG}$ has nonzero value only if $T>T_{90}$. 
In our fiducial case, we set the temporal and spectral energy indices $p_t$ and $p_E$ (i.e., $F_\nu \propto T^{p_t} \nu^{p_E}$) to $-1.5$ and $-1.0$, respectively \citep{Ghisellini10}.

\subsection{EBL attenuation}
\label{subsec:EBL}


We need to take into account the attenuation of gamma-ray photons by the EBL. 
In our fiducial case, we adopt the EBL model of \citet{Razzaque2009}. 
This model is applicable for $z<5$, so that only bursts with $z<5$ are considered in our simulation. 
As we see later in Figure \ref{fig:z_fid_integral}, 
higher-redshift ($z>5$) bursts are not expected to make a significant contribution to the detection rate with CTA \citep[see also][]{Souza2011}. 
For comparison, we also make calculations with the ``Best Fit 06'' EBL model\footnote{http://www.desy.de/\~{}kneiske/downloads.html} of \citet{Kneiske2004} at $z<5$. 
In this model, the gamma-ray horizon, the location where the $\gamma\gamma$ optical depth $\tau_{\gamma\gamma}(E,z)=1$ lies at lower-redshift by a factor of $\sim$~1--2 at $E<100$~GeV compared to our fiducial EBL model. 
However, we found that the resulting differences in the detection rate and the redshift distribution for these two models is not large. 


\section{Modeling of GRB Observations with CTA/LSTs}
\label{sec:GRB_obs_cond_rate}


In this section, we model the CTA observation of GRBs to investigate the GRB detection rate and the expected GRB properties.
The arrays for CTA will be constructed at two sites, each in the northern and southern hemispheres, and designed to have three types of telescopes: LSTs, MSTs, and SSTs. 
The LSTs will be the most crucial component for detecting GRBs because they dominate sensitivity below 200--300~GeV~\citep{CTA11} where EBL attenuation is expected to be less severe, and have the capability of fast slewing (180$^\circ$ in 20~sec). 
In this paper, we consider only one array site and the sensitivity of LSTs alone unless it is explicitly stated otherwise. 

We consider the situation where GRBs are detected by GBM and then are followed up by the LSTs, pointing to the centroid of the GBM error circle. 
Bursts that occur by chance in the LST FOV are not considered because their detection rate is expected to be very low\footnote{Assuming the all-sky GRB rate at the Earth of 10$^3$~yr$^{-1}$, duty cycle of 10~\%, the LST FOV of $4.6^\circ$, and the fraction of the detectable events in the FOV of 10~\%, we get 0.004~yr$^{-1}$. }. 

\subsection{GBM localization}
\label{subsec:GBMlocalization}

Follow-up observations are made only for sufficiently well-localized bursts.
Therefore, we set a threshold value of the localization accuracy of the alerted burst 
for the follow-up observation.
In the case of the MAGIC telescope, its FOV is 3.5$^\circ$, and the threshold of the error radius to start the follow-up is $1.5^\circ$~\citep{Garczarczyk2009}. 
In our simulation for CTA/LSTs, we set the threshold value, $\bf \sigma_{\rm th}$, to $3.5^\circ$ (i.e., the observation is made only when the alerted error radius with 1-sigma accuracy is less than $\sigma_{\rm th}$). 
The error radius of each sample is given as a function of energy fluence using the GBM burst 
data set, where the 1-sigma error radius of individual bursts is estimated as 
$(\sigma_{\rm stat}^2 + \sigma_{\rm sys}^2)^{1/2}~(\equiv \sigma)$. 
The systematic error $\sigma_{\rm sys}$ of $3^\circ$ is assumed for all the bursts for simplicity \citep{Briggs2009}, and the statistical error $\sigma_{\rm stat}$ is given by ground position data (V.~Connaughton, private communication) which are reported $\gtrsim 1$~min after the burst triggers. 

From these assumptions, the fraction of the triggered bursts for which the error radius is less than $\sigma_{\rm th}$ is $\simeq 21$~\%. 
Note that in most cases, only the flight position data, which are more poorly determined compared to the ground one, are available during the prompt phase. 
Hence, as long as the current localization capability of GBM is considered, the simulated localization efficiency (i.e., 21~\%) is larger than reality, which causes an overestimate of the detection rate of the prompt emission. 
Nevertheless, we adopt this assumption expecting the improvement of the localization speed of GBM and its accuracy by the CTA era. 
For the afterglow phase, the above assumption for GBM localization has little influence. 
Without any improvement of the GBM localization, nontrivial followup strategies that compensate for the limited FOV of the LSTs may be helpful to search for the prompt emission of GBM bursts, such as scanning the error circle over time~\citep{VERITAS2011}, or divergent pointing of the LSTs over the error circle.

The error radius of GBM localization is always larger than the radius corresponding to the LST's FOV of $2.3^\circ$.
Hence in our simulation, we take into account the probability that the burst is in the FOV of LSTs after slewing for each burst alerted by the GBM. 
It is given by integrating the two-dimensional Gaussian distribution with the standard deviation of $\sigma$ from 0 to $2.3^\circ$. 
In our fiducial case (i.e., $\sigma_{\rm th}=3.5^\circ$), the probability is about 20~\%. 
The dependence of the detection rate on $\sigma_{\rm th}$ is shown in Section \ref{subsec:dependence3}. 


\subsection{Delay time and detection conditions of CTA} 
\label{subsec:detect_conditionCTA}

When a sufficiently well localized position of a GRB is obtained, the more rapidly CTA points toward the position, the more chances for GRB detection it has. 
In the case of MAGIC-I, if the distribution of the delay time between the start of observations and the GRB trigger time ($T_{\rm delay}$) is fitted by a log-normal probability density function (PDF), then the PDF multiplied by $T_{\rm delay}$ has a peak at $\tau_{\rm delay}$ of $\sim 160$~sec and a standard deviation $\sigma_{\rm delay}$ of $\sim 0.5$~dex~\citep{Garczarczyk2009,Alb+07}. 
Taking into account the ability of rapid slewing of the LST, we assume that $T_{\rm delay}$ of the LST obeys a log-normal distribution with $\tau_{\rm delay}$ of 100~s and $\sigma_{\rm delay}$ of 0.4~dex in our fiducial case. 
In addition we impose $T_{\rm delay}>20$~sec. 
It may be conservative to use $\tau_{\rm delay} = 100$~sec for LSTs considering that the average slewing time of the telescope (not the delay time) of $\simeq 90$~sec for MAGIC-I~\citep{Garczarczyk2009} is more than 70~sec longer than that for LSTs, $< 20$~sec. 
Simulation results for the other values of $\tau_{\rm delay}$ will be shown in Section~\ref{subsec:dependence1}. 

The last step for the GRB observations is to compare the gamma-ray flux with the sensitivity of CTA. 
Here we define $N_{\gamma}$ as the total photon counts and $N_{\rm bg}$ as the total background counts, where both are obtained by integration over the energy range from $E_{\rm low}$ to 300~GeV for a given observation time. 
The low-energy end of the integration $E_{\rm low}$ is given in the next subsection. 
The upper bound of the integration, 300~GeV, is sufficient for discussing the detection rate with LSTs. 
The exposure time is set to be $T_{90} - T_{\rm delay}$ for the prompt emission and at most 4~hours for the afterglow. 
Following the simple estimates of telescope sensitivity~\citep[e.g.,][]{Aharonian2001,CTA10}, we judge that a burst is detected if all of the following conditions are satisfied:
(1) $N_{\gamma} > N_{\rm min}$, 
(2) $N_{\gamma} > m\sqrt{N_{\rm bg}}$, 
and (3) $N_{\gamma} > e N_{\rm bg}$, 
where $N_{\rm min}$ = 10, $m=5$ and $e=0.05$. 
The condition (1) concerns the minimum number of photons required to create the sufficient Cherenkov radiation. 
The conditions (2) and (3) are about the statistical significance and the systematic error, respectively. 
To evaluate these conditions for each sample, we need the effective area of CTA/LSTs, background spectrum, angular resolution, and $E_{\rm low}$. 
These are given in the next subsection. 

\subsection{Performance of the LSTs} 
\label{subsec:LSTs}

\begin{figure}
\centerline{\includegraphics[width=1\columnwidth]{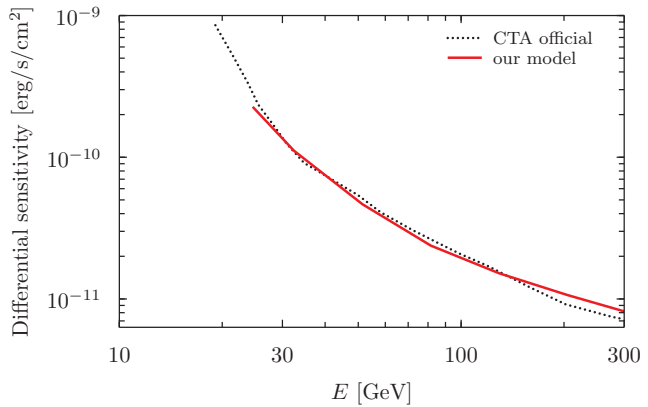}}
\caption{
The differential sensitivity curves of our CTA/LSTs model (red solid curve) and the official CTA (black dotted curve)~\citep{CTA10}. 
The official one includes the contributions from all types of telescopes (i.e., LSTs, MSTs, and SSTs), where at $\lesssim$~200--300~GeV LSTs dominate the sensitivity. 
The exposure time is 0.5~h.
The right end of the horizontal axis (i.e., 300~GeV) is the maximum energy considered in our simulation. }
\label{fig:toy_model,0.5h}
\end{figure}
In this subsection, we construct a toy model for the LSTs, which reproduces the differential sensitivity of CTA that is publicly available \citep{CTA10}. 
In order to calculate the differential sensitivity of our model, we set: the effective area of LSTs after all cuts for gamma rays $A_\gamma$, the effective area for the background $A_{\rm bg}$, the background spectrum $J_{\rm bg}$ in units of particles~s$^{-1}$cm$^{-2}$sr$^{-1}$GeV$^{-1}$, and the angular resolution $\phi$. 
These are given as functions of photon or particle energy.
The effective area $A_{\rm bg}$ is approximately the same as $A_\gamma$ because the background is dominated by electrons at energies below $\sim100$~GeV, inducing electromagnetic showers similar to gamma rays~\citep{CTA10, Aharonian2001}.
Then we assume $A_{\rm bg} \propto A_\gamma$ and $J_{\rm bg} \propto E^{-3.045}$ according to the {\it Fermi}/LAT and the HESS observations of cosmic-ray electrons and positrons~\citep{Fan2010,Aha+08,Abd+09e}. 
In addition, assuming $\phi \propto E^{-0.5}$~\citep{CTA10}, we can calculate the differential count rate of the background as $dR_{\rm bg}/dE =\pi J_{\rm bg} A_{\rm bg} \phi^2 \propto E^{-4.045} A_\gamma$. 
The normalization of $dR_{\rm bg}/dE$ is given below. 

Once the functional form of $A_\gamma$ is given, a shape of the differential sensitivity curve (that is, its energy dependence) is determined. 
We found that by using the effective area of the MAGIC telescope at its trigger level as $A_\gamma$~\citep{MAGIC2011}
\footnote{The effective area at the trigger level decreases toward the low energy more gradually than after all cuts.}, 
we can reproduce well the shape of the official sensitivity curve of CTA at $\lesssim$~200--300~GeV, where the LSTs dominate the sensitivity. 
Therefore as a functional form of $A_\gamma$ at zenith angle $\theta_{\rm{zen}}=20^\circ$, we adopt the MAGIC effective area at the trigger level in our simulation. 
We normalize $A_\gamma$ introducing a factor of $f_A$, where $f_A=1$ means that $A_\gamma$ corresponds to the effective area at the trigger level of MAGIC. 

For any value of $f_A$, we can fit the sensitivity curve of our model to the public one with the normalization of $dR_{\rm bg}/dE$.
In Figure~\ref{fig:toy_model,0.5h}, the official differential sensitivity of CTA is shown as the black dotted curve for the exposure time of 0.5~h, where contributions from all types of telescopes (i.e., LSTs, MSTs, and SSTs) are included, though LSTs dominate the sensitivity for $E\lesssim$~200--300~GeV.
Because the sensitivity for this exposure time is limited by the statistical significance (see the detection condition (2) described in the last part of Section \ref{subsec:detect_conditionCTA}) at $<300$~GeV, we can determine the normalization of $dR_{\rm bg}/dE$ as a function of $f_A$.
In the case of $f_A=1$, we need $R_{\rm bg}\simeq 0.34$~Hz for fitting, where $R_{\rm bg}$ is calculated by the integration of $dR_{\rm bg}/dE$ from 20~GeV to 20~TeV.
The sensitivity curve of our model at $\theta_{\rm zen}=20^\circ$ is shown as the red solid curve in Figure~\ref{fig:toy_model,0.5h}.
One can see that the two curves are similar.
This figure shows the photon energy up to 300~GeV since we calculate the photon counts below this energy in our simulation.
Indeed, the GRB spectrum above $\sim$~100~GeV is expected to be severely attenuated by the EBL, so that our artificial cutoff at this energy does not affect the rate estimate.
Above 200--300~GeV, the MSTs and the SSTs have better sensitivity compared to the LSTs, so that the sensitivity of our model deviates from the total array sensitivity.
In order to keep the two sensitivity curves consistent with each other for arbitrary value of $f_A$, we need $R_{\rm bg} = 0.34f_A^2$~Hz to normalize $dR_{\rm bg}/dE$ considering the detection condition (2).
Although the value of $f_A$ influences the condition (1) and (3), it varies the expected detection rate only a little (see Section \ref{subsec:dependence2}).
In our fiducial case, we assume $f_A=1$.

As $\theta_{\rm zen}$ increases, the shower maximum height gets higher and Cherenkov light is more absorbed, so that the area of light pool becomes large, while its density becomes low. 
In order to take into account the dependence of the effective area on the zenith angle, we use the following form: 
\begin{equation}
A_\gamma(E,\theta_{\rm{zen}})\,=\,
\frac{1}{(\cos \theta_{\rm{zen}})^\xi}
A_\gamma(E',0^\circ)~~, 
\label{eq:Aeff}
\end{equation}
where $E'\equiv E(\cos \theta_{\rm{zen}})^\zeta$, $\xi\,=\,1.7$ and $\zeta\,=\,2.4$~\citep{Aharonian1999}. 
Also for the angular resolution $\phi$, we multiply by a factor of $(E'(\theta_{\rm zen})/E'(20^\circ))^{-1/2}$. 

The differential count rate of gamma rays becomes maximum at $\simeq 60~(\cos \theta_{\rm{zen}})^{-\zeta}$~GeV in our model for a source with its photon index of $-2$, neglecting the EBL attenuation. 
We set $E_{\rm low}$, which is the low-energy end of the integration for the photon counts, at an energy lower than this as 
\begin{equation}
E_{\rm low}~(\theta_{\rm zen}) = 20~(\cos\theta_{\rm{zen}})^{-3.3}~{\rm GeV}~~. 
\label{eq:Elow}
\end{equation}
As we describe in Section \ref{subsec:dependence2}, the exponent of cosine, $-3.3$, has only a small influence on the detection rate. 


\section{GRB Detection Rate with CTA}
\label{sec:grb_rate_results}

In this section, we show the results of our Monte Carlo simulation on the GRB detection rate with the LSTs, which are obtained under the assumptions described in Section \ref{sec:GRB_properties} and Section \ref{sec:GRB_obs_cond_rate}. 
Following is a summary of the simulation process. 
First, in order to generate intrinsic GRB samples, we randomly give $L_{\rm p}$, $z$, $\alpha$ and $\beta$ according to the distributions described in Section \ref{sebsec:prompt}. 
Using several correlations with respect to the prompt emission properties, we determine $E_{\rm p}$, $E_{\rm iso}$, $L_{\rm ave}$, and $T_{90}$ from $L_{\rm p}$; 
for the first two parameters, the deviation from the best fit line of the correlation is given by a Gaussian random variable. 
Then, we can calculate the prompt and afterglow gamma-ray flux arriving at the Earth for each generated burst taking into account the EBL attenuation. 
Second, we set the trigger and the localization conditions to determine the detected and localized events. 
The former is given in terms of the peak photon flux, while the latter is given by the probability of sufficient localization as a function of the fluence (deduced in Section \ref{subsec:GBMlocalization}). 
Then it is judged whether each sample is localized sufficiently well and whether it is in the FOV of the LSTs after slewing to the best position. 
Finally, at the stage of follow-up observation by LSTs, $T_{\rm delay}$ is given by a log-normal distribution, and $\theta_{\rm zen}$ is isotropically distributed independently of $T_{\rm delay}$, where we assume a $10$~\% duty cycle and $\theta_{\rm zen}<60^\circ$ as the observational criteria. 
We evaluate the detection conditions described in Section \ref{subsec:detect_conditionCTA} on each sample taking into account the zenith angle dependence of the array performance (as shown in Section \ref{subsec:LSTs}). 

In this simulation, we consider only one array site and the GBM alerts (with its trigger rate of $250$~yr$^{-1}$) alone. 
We limit our simulation to the GBM bursts with $T_{90}>2$~sec and $z<5$, the fraction of which is $\sim 80$~\% of all GBM bursts. 
Note that this criterion does not greatly affect our final result of the detection rate. 
One of the reasons is that the trigger rate of short GRBs ($T_{90}<2$~sec) and high-redshift GRBs ($z>5$) is typically smaller than the total trigger rate.  
Second, it is nearly impossible to start followup observations of short GRBs within their duration of prompt emission since it takes more than 2~sec to receive the positional information from satellites. 
In addition, short bursts have smaller fluences than the long bursts, which will make the localization of the short GRBs with GBM more difficult.
Finally, detection rate of the high-redshift GRBs is expected to be smaller (especially for afterglows) due to the severe EBL attenuation of gamma rays. 


Hereafter, we classify generated GRB samples into several sets, in accordance with steps from the trigger by GBM to the detection by CTA: 
\begin{description}
\item [{\it Alert}]: 
  GRB samples that are detected by GBM and satisfy $T_{90}>2$~sec and $z<5$. 
\item [{\it CTAobs}]: 
  Samples belonging to {\it Alert} whose positions are determined within zenith angle $\theta_{\rm zen}<60^\circ$ with the error radius $\sigma$ smaller than $\sigma_{\rm th}$. 
  Moreover, the CTA duty cycle of 10~\% is also taken into account to select the samples of this class. 
\item [{\it Pobs}]: 
  Samples belonging to {\it CTAobs} which satisfy the criterion $T_{\rm delay}<T_{90}$. 
\item [{\it Pdet}]: 
  Samples belonging to {\it Pobs} which are actually located within the FOV of LSTs and their prompt emissions are detectable by LSTs. 
\item [{\it Adet}]: 
  Samples belonging to {\it CTAobs} which are actually located within the FOV of LSTs and their afterglows are detectable by LSTs. 
\end{description}
LSTs makes observations for the {\it CTAobs} bursts. 


\subsection{Detection rate in the fiducial case} 
\label{subsec:fiducial-case}

First, we show results of the detection rate with LSTs for our fiducial case in which we assume: 
(1)~only the Band component for the prompt spectrum, i.e., no extra hard component ($R_{\rm extra}=0$), 
(2)~GRBs whose high-energy photon index of the Band component is $\beta < -2$,
(3)~for the afterglow emission, the temporal index $p_t$ of $-1.5$ and the spectral energy index $p_E$ of $-1$,
(4)~the EBL model provided by \citet{Razzaque2009},
(5)~the typical delay time of starting observation, $\tau_{\rm delay}$, of 100~sec and $\sigma_{\rm delay}$ of 0.4~dex,
and (6)~the threshold value of the error radius $\sigma_{\rm th}$ of $3.5^\circ$ to start the follow-up observations. 
The dependence of the detection rate on parameters related to the LST performance and GRB emission is quantitatively discussed in later subsections. 

\subsubsection{Prompt emission} 

With the above assumptions, we obtain 0.03~yr$^{-1}$ as the detection rate of the GRB prompt emission with LSTs (see the first line in Table~\ref{table:GRBrate_theory1}). 
Factors reducing the rate of {\it Alert} to that of {\it Pdet} are as follows. 
Among the samples belonging to {\it Alert} ($\sim 200$~yr$^{-1}$), 21~\% (i.e., 43~yr$^{-1}$) are localized enough to be followed-up by LSTs (see Section \ref{subsec:GBMlocalization}). 
Moreover, imposing $\theta_{\rm zen}<60^\circ$ and duty cycle of 0.1 we get 1.1~yr$^{-1}$ for {\it CTAobs} sample. 
Of these, {\it Pobs} GRBs are 0.45~yr$^{-1}$, and 6~\% of {\it Pobs} GRBs belong to {\it Pdet}. 
Note that only one array site and only the alerts from GBM are assumed in this calculation. 
We discuss more general cases in Section \ref{sec:summary}. 

\begin{figure}
\centerline{\includegraphics[width=0.9\columnwidth]{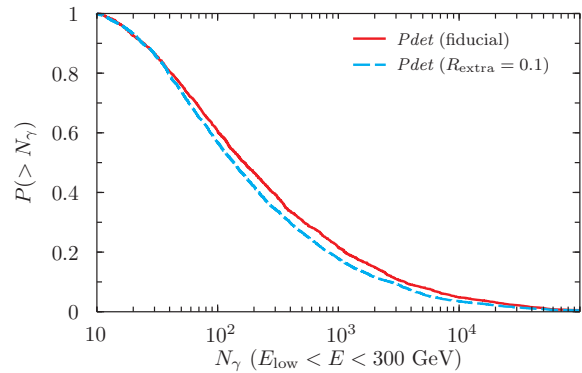}}
\caption{
The cumulative distributions of the total photon counts $N_{\gamma}$, $P(>N_\gamma)$, where $N_{\gamma}$ is calculated by integration over $E_{\rm low} < E < 300$~GeV. 
The low-energy end of the integration $E_{\rm low}$ is defined by Eq. (\ref{eq:Elow}). 
The red solid curve shows {\it Pdet} samples for the fiducial case (see Section \ref{sec:grb_rate_results} or Table \ref{table:GRBrate_theory1} for explanation of classification for our samples). 
The blue dashed curve represents the case of $R_{\rm extra}=0.1$ for {\it Pdet} samples with the other parameters fixed to the fiducial values. 
}
\label{fig:Ngam_fid_integral}
\end{figure}

In Figure~\ref{fig:Ngam_fid_integral}, we show the cumulative distributions $P(>N_\gamma)$ of the total photon counts $N_{\gamma}$, where $N_{\gamma}$ is calculated by integration over $E_{\rm low} < E < 300$~GeV. 
The red solid curve shows {\it Pdet} samples for the fiducial case. 
Once the prompt emission is successfully observed, we can expect the photon counts $N_{\gamma} > 10^2$ with the probability of 60~\%. 
We also calculated the photon counts ($>1$~GeV) expected for {\it Fermi}/LAT, and compared  with $N_{\gamma}$. 
Then it is found that CTA/LSTs can detect 10--100 times more GeV photons than LAT for almost all the {\it Pdet} samples. 
Therefore CTA can provide the temporal and the spectral structure of high-energy emission of GRBs with higher significance than any other current instruments. 

\begin{figure}
\centerline{\includegraphics[width=1.0\columnwidth]{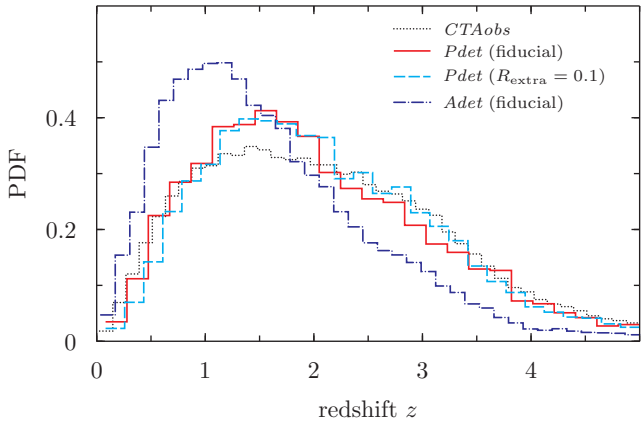}}
\caption{
The simulated redshift distributions are shown as the form of the probability density function (PDF). 
For {\it Pdet} samples (see Section \ref{sec:grb_rate_results}), the red solid line describes the fiducial case, while the blue dash-dotted line is for the case of $R_{\rm extra}=0.1$ with the other parameters fixed to the fiducial values. 
The distribution of {\it Adet} samples and {\it CTAobs} samples in the fiducial case are represented by the blue dash-dotted line and the black dotted line, respectively. 
}
\label{fig:z_fid_integral}
\end{figure}

Figure~\ref{fig:z_fid_integral} shows the redshift distribution of {\it Pdet} samples for the fiducial case with the red solid line in the form of the PDF. 
We found that for {\it Pdet}, $z \sim 1.5$ is typically expected and 90~\% have redshifts less than 3.5. 
These results are little influenced by our limitation of $z<5$, while it can depend on the assumed EBL model. 
Again we find that the difference in the redshift distribution between the model of \citet{Kneiske2004} and that of \citet{Razzaque2009} is very small (see also Table~\ref{table:GRBrate_theory2} with regard to the detection rate). 
For comparison, we show the result for {\it CTAobs} samples (that have not been selected by the CTA performance) with black dotted line. 
We can see that the distribution of {\it CTAobs} is close to that of {\it Pdet}, which implies that the redshift distribution of {\it Pdet} events is mainly determined by the GBM sensitivity. 


\subsubsection{Afterglows} 

The detection rate of the afterglow is obtained as 0.13~yr$^{-1}$. 
Among {\it CTAobs} samples, 12~\% satisfy detection criteria. 
Detections of the afterglow are expected for $\simeq 84$~\% of GRBs whose prompt emission is detectable. 
Table~\ref{table:GRBrate_theory1} summarizes the results for our fiducial model assumptions. 
A similar estimate for the {\it SVOM} satellite will be presented in Section \ref{subsec:rate_svom}. 
\begin{table*}
\caption{
Expected event rates for one array site. 
We separately consider the cases in which each of {\it Fermi}/GBM and {\it SVOM}/ECLAIRs are the alerting detectors. 
Our fiducial model parameters are assumed. 
Each class is defined as follows:
({\it Alert}) Samples of GRBs that are detected by GBM and satisfy $T_{90}>2$~sec and $z<5$; 
({\it CTAobs}) GRB samples belonging to {\it Alert} whose positions are determined within zenith angle $\theta_{\rm zen}<60^\circ$ with an accuracy of $\sigma < 3.5^\circ$. 
  Moreover, we take into account CTA duty cycle of 10~\% to select {\it CTAobs} samples; 
({\it Pobs}) GRB samples belonging to {\it CTAobs} which satisfy the criterion $T_{\rm delay}<T_{90}$; 
({\it Pdet}) GRB samples whose prompt emissions are detectable by LSTs; 
({\it Adet}) GRB samples whose afterglow emissions are detectable by LSTs. 
}
\label{table:GRBrate_theory1}
\begin{center}
\begin{tabular}{lccccc}
                                 & {\it Alert} & {\it CTAobs} & {\it Pobs} & {\it Pdet} & {\it Adet} \\ \hline
 {\it Fermi}/GBM    [yr$^{-1}$]  & $\sim 200$  & 1.1          & 0.45       &      0.03  & 0.13 \\ 
 {\it SVOM}/ECLAIRs [yr$^{-1}$]  & $\sim 56$   & 2.0          & 0.65       &        0.1 & 0.37
\end{tabular}
\end{center}
\end{table*}

In Figure~\ref{fig:z_fid_integral}, blue dash-dotted lines show the redshift distribution of the {\it Adet} samples. 
We can expect typically $z \sim 1$ for {\it Adet}. 
It is found that the peak of the distribution of {\it Adet} samples is shifted toward lower $z$ from the distribution of {\it CTAobs} samples. 
This implies that in contrast to {\it Pdet}, low-$z$ samples are selected by the CTA sensitivity. 
We found that 90~\% of {\it Adet} samples have redshifts less than $2.9$. 


\subsection{Dependence on delay time distribution} 
\label{subsec:dependence1}

\begin{figure}
\centerline{\includegraphics[width=1.0\columnwidth]{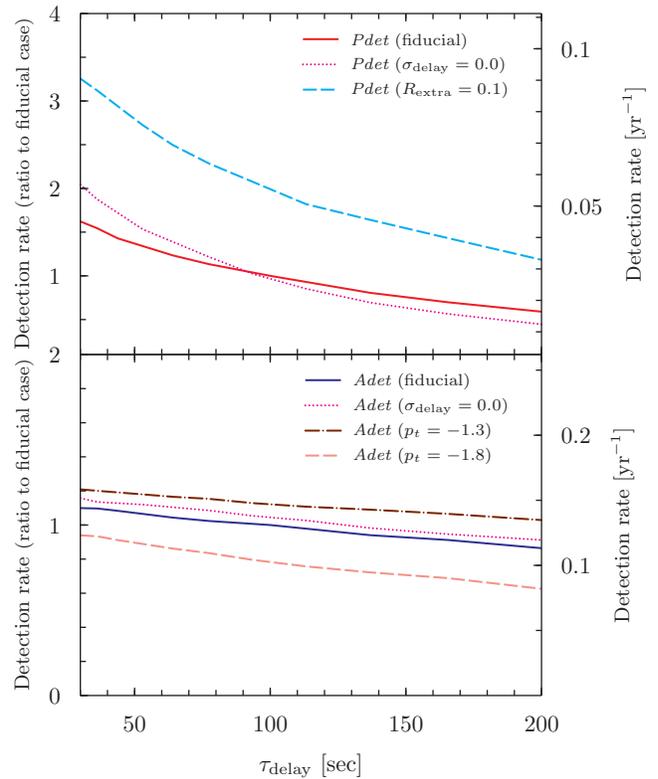}}
\caption{
Dependence of the GRB detection rate on $\tau_{\rm delay}$, where $\tau_{\rm delay}$ is the typical delay time between the follow-up observation and the trigger. 
The right- and the left-hand side of the vertical axis represents the detection rate and its ratio to that for our fiducial case (i.e., $\tau_{\rm delay}=100$~sec), respectively. 
{\it Top panel}: Results for the prompt emission.
The red solid curve  represents the fiducial case except for $\tau_{\rm delay}$. 
The magenta dotted curve represents the case with no variance of the delay time distribution. 
The light-blue dashed curve represents the case in which the extra spectral component with $R_{\rm extra}=0.1$ is introduced.
{\it Bottom panel}: Results for the afterglow emission.
The blue solid curve represents the fiducial case except for $\tau_{\rm delay}$.
The brown dot-dashed and the pink dashed curves represent the cases in which the afterglow temporal index $p_t$ is set to $-1.3$ and $-1.8$, respectively.
}
\label{fig:delay-vs-PAdet}
\end{figure}

In Figure~\ref{fig:delay-vs-PAdet}, we show the dependence of the GRB detection rate on the typical delay time, $\tau_{\rm delay}$, which is introduced in Section \ref{subsec:detect_conditionCTA}. 
The top and the bottom panels represent cases for the prompt emission and the afterglow, respectively.  
In both panels, the horizontal axis represents $\tau_{\rm delay}$, whereas the vertical axis shows the ratio of the detection rate to that for the fiducial parameter set.
Hence, in both panels, the curves labeled as fiducial have the ratio of 1 if $\tau_{\rm delay}=100$~sec. 

First, let us consider the prompt emission (the top panel of Figure~\ref{fig:delay-vs-PAdet}).  
The red solid curve represents the fiducial case except for $\tau_{\rm delay}$. 
If $\tau_{\rm delay}=60$~sec for LSTs, the detection rate is enhanced by a factor of $1.3$. 
The light-blue dashed curve shows the result for the case where the extra component with $R_{\rm extra}=0.1$ is added to a spectrum of the prompt emission in all GRB samples. 
In this case, independently of $\tau_{\rm delay}$, the detection rate is doubled compared to our fiducial case ($R_{\rm extra}=0$). 
In addition, to see the influence of the dispersion of $T_{\rm delay}$ distribution we draw the magenta dotted curve for the extreme case of no dispersion, i.e., $T_{\rm delay}=\tau_{\rm delay}$ for all events. 
At $\tau_{\rm delay} = 90$~sec where it is comparable to the peak of the $T_{90}$ duration distribution of {\it CTAobs} GRBs, the two lines cross each other.
If $\tau_{\rm delay} \gtrsim 90$~sec, the dispersion makes the events with $T_{\rm delay}$ smaller than the central value $\tau_{\rm delay}$, which enhances the detection rate.
On the other hand, if $\tau_{\rm delay} \lesssim 90$~sec, the dispersion makes the events with $T_{\rm delay}$ larger than the central value $\tau_{\rm delay}$, which reduces the detection rate.

Next, let us consider the afterglow (the bottom panel of Figure~\ref{fig:delay-vs-PAdet}). 
As the red solid curve in the top panel, the blue solid curve is for the fiducial case except for $\tau_{\rm delay}$. 
This curve is for the afterglow temporal index $p_t$ of $-1.5$, while the brown dot-dashed and the pink dashed curves are for $p_t=-1.3$ and $-1.8$, respectively.  
These suggest that for the afterglow detection, the important factor is not $T_{\rm delay}$ but others such as the well-localized alert rate and the low-energy sensitivity. 

\begin{figure}
\centerline{\includegraphics[width=1.0\columnwidth]{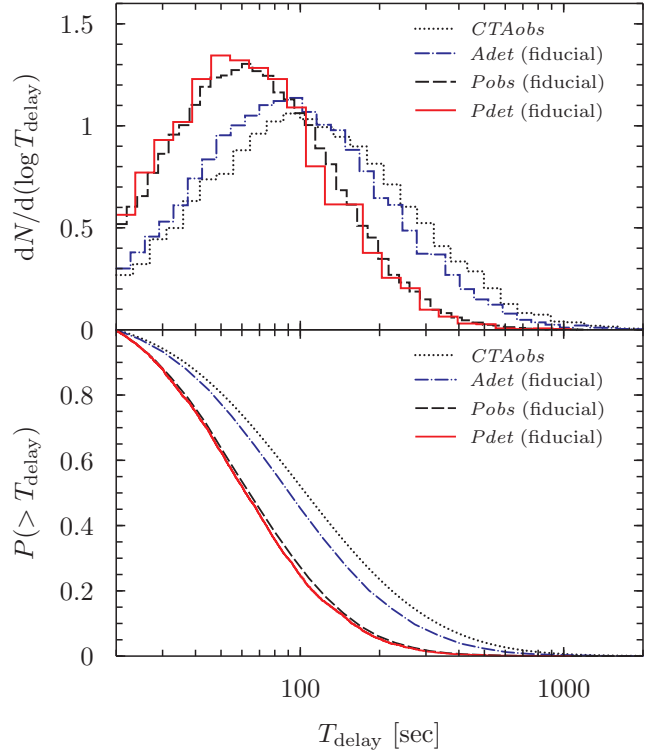}}
\caption{
The delay time distributions for each {\it class} in the case of our fiducial parameter set (see Section \ref{sec:grb_rate_results} or Table \ref{table:GRBrate_theory1} for explanation of classification for our samples).
The bottom panel and the top panel are in the form of the cumulative distributions and ${\rm d}N/{\rm d}(\log T_{\rm delay})$ normalized as $N=1$, respectively. 
Each line represents the different {\it class}: for {\it CTAobs}, the black dotted line; for {\it Pobs}, the black dashed line; for {\it Pdet}, the red solid line; and for {\it Adet}, the blue dot-dashed line. 
}
\label{fig:delay-dist}
\end{figure}

The detection rate of the prompt emission is more sensitive to $\tau_{\rm delay}$ than that of the afterglow as shown in Figure~\ref{fig:delay-vs-PAdet}. 
This simply comes from the fact that $T_{\rm delay}$ affects the number of {\it Pobs} bursts that satisfy $T_{\rm delay}<T_{90}$. 
To see this explicitly, we show in Figure~\ref{fig:delay-dist} the $T_{\rm delay}$ distributions of {\it CTAobs}, {\it Pobs}, {\it Pdet} and {\it Adet} in our fiducial case. 
We can see that the $T_{\rm delay}$ distribution of {\it Pdet} is about the same as that of {\it Pobs}, and they are shifted toward the shorter $T_{\rm delay}$ from that of {\it CTAobs}. 
On the other hand, the distribution of {\it Adet} is only slightly off set from that of {\it CTAobs}. 
In the afterglow case the condition $T_{\rm delay}<T_{90}$ is not required. 

Although shortening $\tau_{\rm delay}$ to less than 100~sec does not have much impact on the improvement of the detection rate, we should note that shorter delay time is necessary for detecting the prompt emission. 
Because the minimum cutoff of $T_{\rm delay}$ is fixed to be 20~sec in our simulation, even when larger $\tau_{\rm delay}$ is assumed, shorter $T_{\rm delay}$ plays a significant role as mentioned above in the explanation of $\sigma_{\rm delay} = 0$. 
Actually, in the bottom panel of Figure~\ref{fig:delay-dist}, for $\tau_{\rm delay}=100$~sec, we can see that $T_{\rm delay}$ is distributed at less than $\sim 100$~sec in most cases where the prompt emission is detectable. 
Therefore it is crucial to keep the delay time as short as possible if the duration of the bursts in the CTA band and in the GBM band are similar to each other. 
In order to reduce $T_{\rm delay}$, fast alerts with good localization is equally important to a rapid slewing of LSTs. 
As an example, we simulate the case in which $T_{\rm delay}$ can not be shorter than 100~sec with the other parameters fixed to the fiducial values, and find that the detection rate of the prompt emission decreases by a factor of 2 in this case. 

\subsection{Dependence on the criterion of the error radius to start follow-up observations} 
\label{subsec:dependence3}

\begin{figure}
\centerline{\includegraphics[width=1.0\columnwidth]{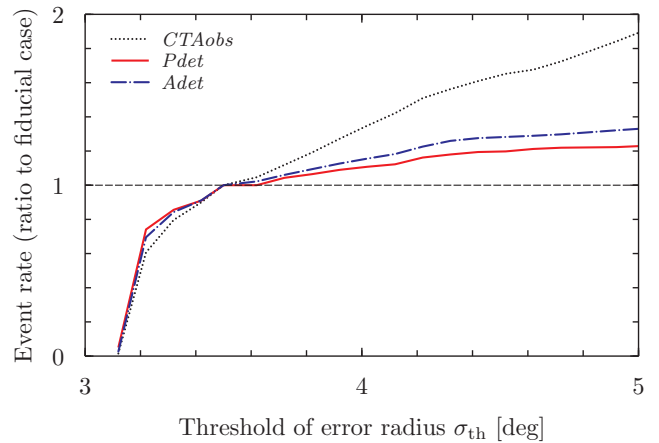}}
\caption{
Dependence of the event rate on the threshold error radius $\sigma_{\rm th}$, which is one of the criteria for LSTs to start the follow-up observations introduced in Section \ref{subsec:GBMlocalization}. 
The results are shown as the ratio to our fiducial case. 
The red solid line represents the result for {\it Pdet}, the blue dot-dashed line for {\it Adet}, and the black dotted line for {\it CTAobs} (see Section \ref{sec:grb_rate_results} or Table \ref{table:GRBrate_theory1} for explanation of classification for our samples). 
}
\label{fig:Errth-vs-PAdet}
\end{figure}

One of the criteria for LSTs to start follow-up observations concerns $\sigma_{\rm th}$, whose fiducial value is set to $3.5^\circ$ (see Section~\ref{subsec:GBMlocalization}). 
Most of the GBM alerts have the error radius larger than this fiducial value. 
The larger $\sigma_{\rm th}$ we use, the more chances of the follow-ups we have, while the efficiency of the detection decreases. 
This is because the bursts with larger error radii are less probable to lie in the FOV of the LSTs. 

In Figure \ref{fig:Errth-vs-PAdet}, we show the dependence of the detection rate on $\sigma_{\rm th}$. 
The results are shown as the ratio to the fiducial case. 
The red solid line ({\it Pdet}) and the blue dot-dashed line ({\it Adet}) show the increase in the detection rate by a factor of 1.2--1.3 for $\sigma_{\rm th} = 5^\circ$ compared to $3.5^\circ$. 
On the other hand, one should keep in mind that the more rapid increase of {\it CTAobs} (black dotted line) shows the decline in the detection efficiency (i.e., the ratio of {\it Pdet} or {\it Adet} to {\it CTAobs}) by a factor of 0.6--0.7. 
On the contrary, if we take $\sigma_{\rm th}$ to be less than $3.2^\circ$, most of the GBM alerts do not satisfy this criterion, so that the detection rate decreases very rapidly. 

It can be seen from Figure \ref{fig:Errth-vs-PAdet} that the detection rates of {\it Pdet} and {\it Adet} saturate around $\sigma_{\rm th} = 5^\circ$, so that it seems better to set $\sigma_{\rm th}$ to near this value as long as we take a simple strategy of the follow-up observations, i.e. all 4 LSTs point toward the centroid of the GBM error circle. 

\subsection{Dependence on other parameters} 
\label{subsec:dependence2}

\begin{table*}
\caption{
Parameter dependence of the detection rate for one array site of CTA/LSTs, where only GBM is assumed as the alerting detector. 
The cases other than the fiducial one are for those in which one of the model parameters takes different value from the fiducial parameter set (with the other parameters fixed). 
Note that the localization accuracy of GBM is estimated optimistically for the prompt emission phase compared to the current localization accuracy. 
See Section \ref{subsec:dependence2} for the explanations of each {\sf Case}. 
}
\label{table:GRBrate_theory2}
\begin{center}
\small
\begin{tabular}{lccccccccc}
                       & fiducial   & {\sf Case~1}             & {\sf Case~2} & {\sf Case~3}        & {\sf Case~4} & {\sf Case~5} & {\sf Case~6} & {\sf Case~7} & {\sf Case~8} \\
                       &            & $\tau_{\rm delay}=60$~s  & $\beta > -2$ & $R_{\rm extra}=0.1$ & Kneiske      & $p_t=-1.3$   & $p_t=-1.8$   & $p_E=-0.5$   & $p_E=-1.5$ \\ \hline
{\it Pdet} [yr$^{-1}$] & 0.03 & 0.04               & 0.04   & 0.06          & 0.02   & --           & --           & --           & -- \\
{\it Adet} [yr$^{-1}$] & 0.13 & 0.14               & --           & --                  & 0.11   & 0.15   & 0.10   & 0.19   & 0.03
\end{tabular}
\end{center}
\end{table*}

\begin{figure}
\centerline{\includegraphics[width=1.0\columnwidth]{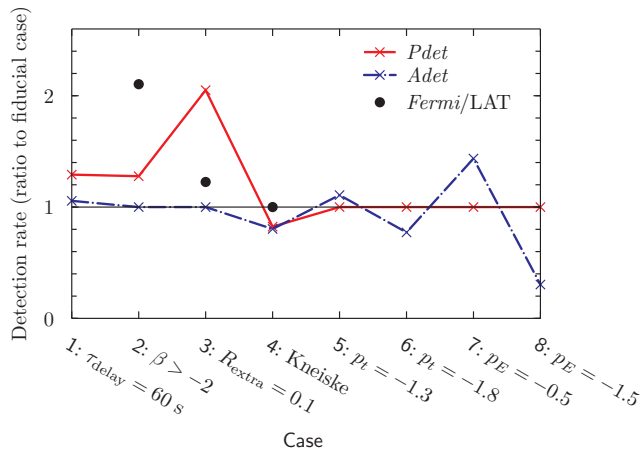}}
\caption{
The results in Table \ref{table:GRBrate_theory2} are plotted as the ratio of the detection rate to the fiducial case.
Each case corresponds to those of the table \ref{table:GRBrate_theory2}.
The red solid and blue dot-dashed lines represent the results for {\it Pdet} and {\it Adet}, respectively.
The same plot but for the simulated LAT detection rate, which is expected as 12~yr$^{-1}$ in our fiducial case, are shown with the filled circle for {\sf Case 2, 3}, and {\sf 4}.}
\label{fig:ratio_modeluncertainty}
\end{figure}

Table~\ref{table:GRBrate_theory2} summarizes how the detection rate changes when one of our model parameters takes different values from that in our fiducial parameter set (with the other parameters fixed). 
Each column describes the result for different cases:
({\it first column}) The fiducial case. 
({\sf Case 1}) The typical delay time of the follow-up $\tau_{\rm delay}$ is set to 60~sec. 
({\sf Case 2}) The case including $\beta>-2$ according to the BATSE observation. 
({\sf Case 3}) The extra spectral component of $R_{\rm extra}=0.1$ is assumed for all the samples. 
({\sf Case 4}) The EBL model by \cite{Kneiske2004} is used. 
({\sf Case 5} and {\sf Case 6}) The temporal index of the afterglow $p_t$ is taken as $-1.3$ and $-1.8$, respectively. 
({\sf Case 7} and {\sf Case 8}) The spectral energy index of the afterglow $p_E$ is taken as $-0.5$ and $-1.5$, respectively.
Figure~\ref{fig:ratio_modeluncertainty} shows the same results presented in Table \ref{table:GRBrate_theory2} but plotted as the ratio to the fiducial case.
The red solid and blue dot-dashed lines represent the result for {\it Pdet} and {\it Adet}, respectively. 

The detection rate for {\sf Case~3} is about twice as large as the fiducial case. 
This fact remains unchanged for different $\tau_{\rm delay}$, which is represented in the top panel of Figure~\ref{fig:delay-vs-PAdet} with the light-blue dashed curve. 
This is because $R_{\rm extra}$ of 0.1 makes the bursts with intrinsically soft Band spectrum with $\beta \lesssim -2.5$ detectable; these soft bursts account for about half of {\it Alert} and {\it CTAobs} samples. 
By contrast, $R_{\rm extra}=0.1$ has little influence on the photon-count distribution and the redshift distribution. 
Comparing {\sf Case~2} with {\sf Case~3}, the latter shows a larger increase of the detection rate compared to the former. 
This simply reflects that the fraction of bursts with $\beta>-2$ is less than those with $\beta \lesssim -2.5$. 

We also calculate the expected LAT detection rate of {\sf Case~2} and {\sf Case~3}, and then obtain 24~yr$^{-1}$ and 14~yr$^{-1}$, respectively. 
The former is more than three times as large as the observed LAT rate for the bursts of $T_{90}>2$~sec and $z<5$ (about 7--8~yr$^{-1}$), which seems unrealistic. 
The calculated LAT event rate for our fiducial case and {\sf Case~3} are similar to each other and about 1.5--2.0 times as large as the observed one. 
These somewhat large rates appear to be consistent with the analysis by \citet{Beniamini2011}, claiming the existence of the spectral softening below the LAT band in some of bright bursts, since we have not taken into account any spectral softening feature in our simulations. 
We discuss this point a bit more in Section \ref{sec:summary}. 

In addition to the cases summarized in Table~\ref{table:GRBrate_theory2}, we studied the dependence of the detection rate on the luminosity function (see Eq. (\ref{eq:rho_z}) and (\ref{eq:phi_L})), the normalization of the effective area, $f_A$ (see Section \ref{subsec:LSTs}), and the low-energy end of the LSTs sensitivity, $E_{\rm low}$ (see Eq. \ref{eq:Elow}). 
For the luminosity function, we found that when one of the 6 parameters included is varied from the best fit value within its errors, the rate changes at most by a factor of $\simeq$~0.8--1.3 for both {\it Pdet} and {\it Adet}. 
For $f_A$, we see its influence on the detection rate by varying the value from $0.3$ to $3$. 
At this time, as a function of $f_A$, the background count rate $R_{\rm bg}$ shifts from about $0.03$~Hz to $3$~Hz to keep the sensitivity of our LSTs model consistent with the official one. 
We found that this range of $f_A$ changes the detection rate by a factor of 0.8--1 for {\it Pdet}, though the change of the {\it Adet} rate is negligible. 
This reflects the fact that for non-detected prompt emissions, the detection conditions (1) and (2) (which are described in the last part of Section \ref{subsec:detect_conditionCTA}), have comparable importance, while for non-detected afterglows, the condition (2) is the most strict. 
Note that the expected photon counts $N_\gamma$ can vary in proportion to $f_A$. 
For $E_{\rm low}$, we see the detection rate for the various exponent of cosine in Eq.(\ref{eq:Elow}). 
By varying it from $-4.3$ to 0, we found that the detection rate changes by a factor of $\simeq$~1.0--1.2 for {\it Pdet} and $\simeq$~0.9--1.1 for {\it Adet}. 
Therefore the luminosity function, $f_A$, or $E_{\rm low}$ is not more sensitive to the detection rate than the other source parameters.
\subsection{CTA detection rate for {\it SVOM} GRBs} 
\label{subsec:rate_svom}

Finally, we roughly estimate the detection rate for the case of alerts from {\it SVOM}/ECLAIRs, whose launch before the CTA operation is being planned~\citep{Schanne2010} and which will provide well-localized alerts ($<10^\prime$) of about 80~yr$^{-1}$~\citep{Paul2011}. 
Here we assume that its duration distribution is the same as the Burst Alert Telescope (BAT) onboard {\it Swift} because the two detectors, ECLAIRs (4--250~keV) and BAT \citep[15--150~keV:][]{Sakamoto2007}, cover a similar energy range.
The {\it Swift} data on $T_{\rm 90}$ was taken from the web site
\footnote{http://swift.gsfc.nasa.gov/docs/swift/swiftsc.html} 
(from GRB~041217 to GRB~110519A).
80~\% of ECLAIRs GRBs are expected at $z<6$~\citep{Paul2011} and $\sim 90$~\% of the BAT bursts are long GRBs. 
Hence we assume that the fraction of all GRBs triggering ECLAIRS that have $T_{90}>2$~sec and $z<5$ is 70~\%. 
We set the delay time $T_{\rm delay}$ of $80$~sec for all bursts for simplicity, while we assumed typically $T_{\rm delay} = 100$~sec for GBM. 
This is because the {\it SVOM} alerts are expected to be faster than the GBM ones (i.e., $< 1$~min~\citep{Schanne2010}). 

With the above assumptions, we can estimate the fraction of well localized events for which $T_{90}>T_{\rm delay}$ is $\simeq33$~\% for long GRBs
\footnote{GBM requires higher fluence to localize bursts, so that the duration of the localized bursts tends to be longer than that of the detected bursts to some extent. This is responsible for smaller ratio of bursts with $T_{90}>T_{\rm delay}$ to the localized bursts for BAT (i.e., 33~\%) than that for GBM (i.e., 43~\%). }. 
It is qualitatively expected that the fraction of {\it Pdet} or {\it Adet} to the bursts in the LST FOV localized by ECLAIRs is lower than that by GBM. 
The reasons for this are as follows: 
(1)~ECLAIRs does not need high fluence for localization as much as GBM, which leads to a lot of burst samples with dim flux in the CTA band.
(2)~ECLAIRs has better sensitivity for softer bursts rather than the hard ones, which again leads to the small number of events detectable with CTA. 
Also, the redshift distribution of bursts localized by ECLAIRs shifts to higher redshift than that by GBM, so that larger fraction of events are severely affected by the EBL attenuation. 
In evaluating the fraction of {\it Pdet} or {\it Adet} to the bursts in LST FOV quantitatively, we do not simulate in detail the follow-up observation of ECLAIRs bursts with CTA taking into account its performance, but we just assume that the efficiency is the same as the case of {\it Swift}/BAT -- if BAT is assumed to be the burst trigger and its trigger threshold for peak photon flux is simply set to 0.4~ph~s$^{-1}$cm$^{-2}$ in the 15--150~keV band (about 90~\% of actually detected BAT bursts have their peak flux above this value), we obtain lower detection efficiency than the case of GBM by a factor of 0.54 for the prompt emission and 0.34 for the afterglow in the case of $\sigma_{\rm th} = 3.5^\circ$, $\tau_{\rm delay}=80$~sec for BAT, and 100~sec for GBM. 
Additionally, we must consider that SVOM operations feature a bias of preferentially pointing toward the anti-solar direction, which increases the probability of follow-up observations with ground-based telescopes. 
We set the increase of the detection rate by a factor of 1.4 referring to~\cite{Gilmore2010}. 

With the above assumptions we can estimate the CTA detection rate for alerts from ECLAIRs as $\simeq 0.1$~yr$^{-1}$ for the prompt emission, while $\simeq 0.37$~yr$^{-1}$ for the afterglow (see Table~\ref{table:GRBrate_theory1}), which is about three times higher rate than GBM. 
Hence {\it SVOM} will probably become better for the detection of GRBs with CTA than {\it Fermi}. 
Note that the GBM localization accuracy is estimated optimistically for the prompt emission phase (see Section \ref{subsec:GBMlocalization}), and also note that, unless the SVOM alerts are transmitted to enough ground stations, the delay time of 80~sec for SVOM alerts turns out to be an optimistic assumption. 


\section{Summary and Discussions}
\label{sec:summary}

In this paper, we have presented the prospects for detection of GRBs with the LSTs of CTA, the most vital component for GRB observations with their fast slewing capability and the best sensitivity at the lowest energies.
We have modeled and simulated the follow-up observation of GRBs with $T_{90}>2$~sec and $z<5$ alerted by {\it Fermi}/GBM. 
Our GRB population is modelled according to a given luminosity function that is consistent with Swift observations together with well-known spectral correlations. 
We note the following strengths of our model: 
(1)~It reproduces the fluence and duration distributions of GBM bursts (Figure~\ref{fig:S_vs_T90,S-pdf}). 
(2)~The differential sensitivity of our model LSTs is consistent with the official one given by the \citet{CTA10} in the energy range less than a few 100~GeV (Figure~\ref{fig:toy_model,0.5h}). 
(3)~The {\it Fermi}/LAT detection rate is predicted to be 12~yr$^{-1}$ for the bursts satisfying both $T_{90}>2$~sec and $z<5$, which is roughly consistent with the actual rate of 7--8~yr$^{-1}$. 

Assuming {\it Fermi}/GBM alerts alone, our fiducial parameter set predicts the GRB detection rate with LSTs for one array site (i.e., north or south) as 0.03~yr$^{-1}$ for the prompt emission and 0.13~yr$^{-1}$ for the afterglow emission (Table~\ref{table:GRBrate_theory1}). 
The expected event rates become larger when two array sites of CTA and additional alerts from {\it SVOM}/ECLAIRs are taken into account --- if the array performance for the two sites are the same in the energy range less than a few 100~GeV, the detection rates go up to about 0.3~yr$^{-1}$ and 1~yr$^{-1}$ for the prompt and afterglow emissions, respectively, where we assume no overlap of FOVs of GBM and ECLAIRs. 
Note that in the above estimates, we optimistically estimate the onboard-localization ability of GBM in the prompt phase, expecting improvement by the CTA era (see Section \ref{subsec:GBMlocalization}). 
For the afterglow emission, our treatment of the GBM localization ability hardly affects the rate estimation.

For our fiducial assumptions,  once CTA succeeds in detecting the prompt emissions for GBM alerts, the total photon counts $N_\gamma$ are expected to be $> 10^2$ for 60~\% of CTA detected events. 
Our simulation also shows that $N_\gamma$ is 10--100 times larger than the number of GeV photons expected by LAT for the same bursts. 
This suggests that CTA can obtain the time resolved light curve in the energy range greater than a few tens of GeV with higher statistics than ever before. 
We expect 90~\% of the prompt burst detected by CTA to have redshifts less than 3.5, and 90~\% of the afterglows to have less than 2.9. 
Because of the follow-up observation after the GBM alerts, the redshift distribution is more affected by the GBM sensitivity rather than that of CTA. 
Hence, more frequent detections of high-$z$ prompt emission are expected for {\it SVOM} alerts. 

Studying the dependence of the detection rate on our model parameters, we found the following results: 
\begin{description}
\item[(i)]
For the prompt emission, if all GRB samples have the extra power-law component with luminosity 10~\% of the Band component, the detection rate with CTA increases by a factor of 1.9. 
\item[(ii)]
For the afterglow, the spectral index has a relatively large effect on the detection rate. 
It decreases by a factor of $3$ with the extreme assumption that all bursts have the softer spectral energy index of $-1.5$, while in the harder case of $-0.5$, the rate increases by a factor of 1.5. 
\item[(iii)]
If the typical value of the delay time between the start of observations and the GRB trigger time is 1~min, the detection rate of the prompt emission increases by a factor of about 1.3 compared to our fiducial case. 
On the other hand, the rate of the afterglow depends slightly on the delay time. 
Although the shorter delay time does not have much effect on the enhancement of the detection rate itself, the ability of fast slewing ($\lesssim 100$~sec) is necessary for catching the prompt emission. 
\item[(iv)] 
It seems better to make follow-up observations for the alerts with the error radius of up to $\sim 5^\circ$, as long as all 4 LSTs simply point toward the best position localized by GBM, although the detection efficiency is decreased. 
\item[(v)] 
The detection rate varies by less than 30~\% for variation of the other parameters including those of the luminosity function. 
\end{description}

Note that the LAT detection rate is estimated to be 14~yr$^{-1}$ in the case of (i), which is about 1.8--2.0 times as large as the observed LAT event rate.  
This discrepancy may be due to the existence of the spectral softening below the LAT band, suggested for subset of bright events \citep{Beniamini2011}, our crude (but reasonable) LAT detection conditions, etc. 
Even if the spectral softening effect is a dominant cause, the detection rate with CTA is not necessarily expected to decrease from our results by a similar factor, 1--2, due to the following possible reasons. 
First, the spectral softening below the LAT band have been only suggested for some of bright bursts~\citep{Beniamini2011}.  CTA will be sensitive even for typical (less bright) bursts, which might not have the spectral softening. 
Second, the bursts with the steep Band component below the LAT band may be hardly detected by LAT, whereas they might have an extra hard component at $>10$~GeV, which can be relevant for CTA. 
We have not taken into account such possibilities in our simulations, which remain as future work. 
In any case, CTA may clarify the properties of the extra hard component and spectral softening of the Band component, which would strongly constrain the prompt emission mechanism. 

It is crucial to increase the alert frequency with good localization in order to achieve higher detection rates. 
Hence the improvement of the localization accuracy by future GRB alert facilities is important. 
Let us consider the alerting detector which has the same characteristics (e.g., sensitivity) as GBM except for the localization ability, and simply assume that the alerts from the detector have a constant error radius $\sigma$ for all the alerted bursts. 
In this case, if we can use the localization error $\sigma$ less than the threshold $\sigma_{\rm th}$, the probability of localization in the FOV gets higher, while the frequency of follow-up observations do not vary ({\it CTAobs}~$\sim 5$~yr$^{-1}$). 
The increase of the probability saturates at $\sigma \sim 1^\circ$, and then the detection rate of {\it Pdet} and {\it Adet} for this detector are enhanced to $\simeq 0.2$~yr$^{-1}$ and $\simeq 1.2$~yr$^{-1}$, respectively. 
It implies that future alerting detectors with the better localization ability are more desirable. 

As countermeasures against the alerts with large localization error, two observing strategies are proposed: 
one is to point each LST at different directions in the GBM error circle, while the other is to scan the region instead of pointing one location. 
The optimal strategy for each observation can be determined by the estimate of the detection probability, which requires the information on the trigger (such as the degree of localization error, the brightness, the expected delay time, and the zenith angle). 
In order to perform this and to increase the chance of detection, we need more detailed study on the sensitivity for each strategy as a function of some important parameters (such as zenith angle, the extended FOV, scan speed, etc.), which remains as a future work. 

In the VHE region, the future High Altitude Water Cherenkov (HAWC) Observatory mission \citep{AbeysekaraHAWC2011} may also detect GRBs with its large FOV ($\sim15$~\% of all the sky) and high duty cycle ($\sim100$~\%), while the GRB spectrum may be difficult to determine by HAWC, because of its small effective area and low energy resolution at $\sim30$~GeV compared to CTA \citep{goodmanHAWC2011}.  Hence, CTA and HAWC are complementary and both types of observations are important. 

Very recently, \citet{Gilmore2011a,Gilmore2011b} and \citet{Bouvier2011arXiv} independently studied on expectations for GRB detection rates by CTA, and we find their results are broadly in agreement with our conclusions here. 

\section*{Acknowledgments}

The authors thank R.~Gilmore, A.~Bouvier, and V.~Connaughton for valuable communications and also thank anonymous referee, 
R. Margutti, M. Mori, M. Ohno, and T. Yoshikoshi for helpful comments.
K.~M. acknowledges financial support by a Grant-in-Aid from JSPS and CCAPP.
This work is supported  in part by JSPS Research Fellowships for Young Scientists No.231446 (K.~T.) and in part by grant-in-aid from the Ministry of Education, Culture, Sports, Science, and Technology (MEXT) of Japan No.~22540278 (S.~I.), No.~21740184 (R.~Y.), Nos.~21684014, 22244019, 22244030 (K.~I.).


\bsp

\label{lastpage}

\end{document}